%% file: main.tex
\documentclass[preprint,12pt,authoryear]{elsarticle}

\usepackage{amssymb}
\usepackage{amsmath}
\usepackage{float}
\usepackage{lineno}

\usepackage{subcaption}

\usepackage{siunitx}
\DeclareSIUnit\parsec{pc}

\journal{Astroparticle Physics}

\begin{document}

% \title{Inverse Compton Emission from WIMP Dark Matter at the Galactic Centre}
\begin{frontmatter}

\title{A refined model of Secondary Photon Emission from Heavy WIMP annihilations in the Galactic Centre} %% Article title

\author[IISER,MPIK]{Rajat Shinde}
\author[UIB,MPIK]{Julia Djuvsland} 
\author[MPIK]{Davide Depaoli}
\author[MPIK]{Jim Hinton} 

\affiliation[IISER]{organization={Indian Institute of Science Education and Research-Pune},
            addressline={Dr. Homi Bhabha Road, Pashan},
            city={Pune},
            postcode={411008},
            %state={Pashan},
            country={India}}

\affiliation[MPIK]{organization={Max-Planck-Institut für Kernphysik},
            addressline={Saupfercheckweg 1},
            city={Heidelberg},
            postcode={69117},
            country={Germany}}
            
\affiliation[UIB]{organization={Department of Physics and Technology, University of Bergen},%Department and Organization
            addressline={Allégaten 55}, 
            city={Bergen},
            postcode={5007}, 
            country={Norway}}

%% Abstract
\begin{abstract}
Heavy Weakly Interacting Massive Particles (WIMPs) remain a prominent yet less constrained dark matter (DM) candidate, with the Galactic Centre (GC) serving as a prime target for indirect detection via gamma-ray signals. Extending our previous work that highlighted the significance of secondary inverse Compton (IC) emission from annihilation-produced electrons, we expand the analysis to a broader range of WIMP masses and introduce a more realistic spatially-dependent modelling framework for the GC environment. This approach incorporates complexities such as the three-dimensional DM distribution, spatially varying radiation and magnetic fields, and electron transport mechanisms like Galactic winds and diffusion. We assess the impact of these environmental factors on both the spatial and spectral characteristics of the resulting secondary emissions. Our results demonstrate the robustness and necessity of incorporating this emission, and highlight its role in enhancing the prospects for detecting heavy WIMPs through observations of the inner Galaxy. We provide the resulting data products to the community to support future analyses and observational studies.
\end{abstract}


% \end{highlights}

%% Keywords

\end{frontmatter}

% \linenumbers


\section{Introduction}
\label{sec:intro}
\input{intro}

\section{The Galactic Centre Environment}
\label{sec:gc}
\input{gc}

\section{WIMP spectra}
\label{sec:wimps}
\input{wimps}

\section{3D Simulation Approach}
\label{sec:approach}

\input{approach}

\section{Results}
\label{sec:results}
\input{results}

\section{Discussion}
\label{sec:discussion}
\input{discussion}

\section{Conclusions}
\label{sec:conclusions}
\input{conclusions}

\section*{Acknowledgements}
J.D. is funded by the Research Council of Norway, project number 301718
and grateful for the hospitality of the Non-Thermal Astrophysics division at
MPIK.

R.S. gratefully acknowledges funding support from the Max-Planck-Gesell\-schaft (MPG) – IISER collaboration, as well as the hospitality provided by MPIK.

% %% The Appendices part is started with the command \appendix;
% %% appendix sections are then done as normal sections
% \appendix
% \section{Example Appendix Section}


%%       \citet{<label>} ==> Lamport (1994)
%%       \citep{<label>} ==> (Lamport, 1994)
%%

\bibliographystyle{elsarticle-harv} 
\bibliography{cas-refs}


\end{document}

%% file: intro.tex
Dark Matter (DM) makes up 26\,\%~\citep{PhysRevD.110.030001} of the total energy budget in our universe and there is ample evidence for its existence - e.g. through gravitational lensing observations, the study of the cosmic microwave background (CMB) and other measurements. However, the particle nature of DM 
remains unknown and many competing theories exist.

A prime particle candidate for DM is a thermal relic weakly interacting massive particle~\citep[WIMP, see e.g.][]{Arcadi_2025}. WIMPs are typically assumed to have masses in the GeV range and interaction cross sections typical of that of the weak interaction. This scenario yields rather naturally the observed DM abundance, a fact sometimes referred to as the {\it WIMP Miracle}. Supersymmetry or extra-dimensional theories are extensions of the Standard Model commonly invoked to explain such new particles. Increasing though this paradigm is in tension with tight upper limits from Direct Detection experiments and the non-observation of DM candidates with the LHC~\citep[see for example]{ReviewWIMPs}, increasing interest in relics at somewhat higher masses. 

%which results in the fact that they can describe the currently observed abundance of DM. WIMP candidates can be obtained by extensions of the Standard Model of particle physics and are featured in e.g supersymmetric and extra-dimensional theories.

The paradigm of GeV to TeV mass range WIMPs is testable with gamma ray observations, as gamma rays are expected to be produced during the decay or annihilation of WIMPs. Competitive limits are being set with current day instruments such as Fermi-LAT\citep{Ackermann_2017}, H.E.S.S.~\citep{Abdalla_2022} and HAWC~\citep{albert2023}, and a leap in sensitivity is expected in the next few years when the next-generation gamma ray observatories CTAO~\citep{Acharyya_2021} and SWGO~\citep{Viana_2019} go online. The current limits exclude a thermal relic WIMP up to masses of a few GeV but the limits are computed taking only the \textit{direct photon} emission from DM annihilation into account. We have shown however, that WIMP annihilation in the Galactic Centre (GC) also gives rise to sizeable \textit{indirect photon} emission~\citep{DJUVSLAND2023}. The emission comes from the direct electrons\footnote{The term electrons is used to describe both direct electrons and direct positrons throughout the paper.} that are emitted during the annihilation chain of WIMP dark matter. The direct electrons produce energetic photons via inverse Compton (IC) scattering of ambient photons or as synchrotron radiation and Bremsstrahlung. These indirect photons have typically lower energies than the direct photons but are still part of the WIMP signal and should therefore not be neglected in WIMP studies. Yet, the indirect photon component requires a certain level of understanding of the emission region, including the 3D distribution of the radiation field. Moreover, this indirect signal is expected to be diffuse, as the high-energy electrons are predicted to travel for several hundred parsecs before coming to rest. Constraints on dark matter models that include the inverse Compton component have been presented earlier by \citep{Cirelli_PAMELA} in the context of interpreting the excess of $e^+e^-$ observed in PAMELA \citep{PAMELA}, and very recently in \citep{Rocamora} from employing observational limits from HAWC and LHAASO \citep{LHAASO}. Another study that included the IC component to improve on HAWC upper-limits was presented in \citep{HAWC_DM_IC}. The authors in that work emphasized that future work should incorporate diffusion effects and conduct systematic investigations of the IC component across different propagation scenarios.

With this work, we expand our previous study~\citep{DJUVSLAND2023} by examining a wider range of WIMP masses made accessible through~\citep{HDM}. % in sect ...?
 As in our previous work, we produce results considering WIMP annihilation near the GC into the three channels $W\overline{W
},\, b\overline{b}\, \text{and}\, \tau\overline{\tau}$.
We further improve the model employed for computing secondary emission by constructing a kiloparsec-sized, three-dimensional, spatially dependent approach for electron injection, cooling, and radiation. We study its robustness under changes in several parameters and provide ready-to-use results for the community. %in sect ..?

% \emph{GC...IC (+synchrotron)... utility...DM candidates beyond WIMPs?....}

%% file: gc.tex
The centre of our galaxy is a complex region hosting various gamma-ray emitters. 
At its centre is the supermassive black hole Sgr A*, surrounded by massive star clusters, supernova remnants, pulsars, and pulsar wind nebulae, embedded in dusty molecular clouds. Charged particles emitted in this region are subject to the local electromagnetic fields and the transport properties of the region.   

WIMP models predict the highest WIMP-density at the centre of galaxies, making the centre of our own galaxy a prime target for WIMP searches due to its proximity. Yet, at the bright centre, the WIMP signal can be difficult to isolate from the astrophysical foregrounds. Therefore, typically, the central region is excluded from WIMP searches and only the regions above and below the Galactic disc are considered. Even there, the environment is special and shows a reduced diffusion with respect to the Galactic average~\citep{HESS_2016}. As in the previous work, we assume a Kolmogorov turbulent spectrum. Yet, the transport of relativistic charged particles in this region is expected to be advection dominated over a wide range of energies as a wind~\citep{Bland_Hawthorn_2003, Di_Teodoro_2018} is blowing out of the Central Molecular Zone (CMZ). 

The Galactic magnetic field is complex and can be divided into a disk field and an extended halo field with out-of-plane components~\citep{Jansson_2012}. While the topology is probably even more complex at the centre of the galaxy ($\lesssim\,100\,$pc)~\citep{Guenduez_2020} we employ the model given in~\citep{Jansson_2012} and~\citep{Jansson_2012_2} as a further complication is beyond the scope of this work. The model employed here includes large-scale regular fields, striated fields and small-scale random fields that are fitted to data from Faraday rotation measures and polarised synchrotron radiation.

The radiation field used in this work has two components: the CMB, i.e. thermal photons which permeate the universe with a blackbody temperature of 2.7$^{\circ}$\,C and interstellar radiation field (ISRF) photons from dust and starlight emission. The latter has a spatial dependency and corresponds to the self-consistent model of the broad-band continuum emission of our Galaxy as described in~\citep{Popescu_2017}. The model was derived from modelling maps of all-sky emission in the infrared and submillimetre regimes, and is 2D axisymmetric, resulting in an isotropic IC modelling.

Although galaxies are expected to be embedded in large DM halos, the spatial distribution of DM at the GC is still unclear. While it seems established that the highest concentration of DM coincides with the GC, several competing profiles are considered. The profiles are typically derived from N-body simulations or rotation curves and can be grouped into two categories: Cored profiles, which have a flat DM density at the inner GC region and cuspy profiles, where the DM density peaks at the very centre of the GC. A universal cuspy profile obtained from N-body simulations is the so called NFW profile (\citep{NFW1}):

\begin{equation}
\rho(r) = \frac{\rho_0}{\left(\frac{r}{r_s}\right) \left(1 + \frac{r}{r_s}\right)^2}
\end{equation}

where $\rho_0$ and $r_s$ are the scale density and radius, which vary from halo to halo. Another popular candidate favoured by recent simulations is the Einasto profile~\citep{Einasto_1965}, which can be parametrised as:
\begin{equation}\label{eq: einasto}
    \rho_{\mathrm{E}}(r) = \rho_0 \exp\left(- \frac{2}{\alpha}\left[\left(\frac{r}{r_s}\right)^{\alpha}-1\right]\right).
\end{equation}
 A cored density profile can be obtained from the Einasto profile by assuming a constant density for the innermost part, according to 
\begin{equation}\label{eq: cases}
    \rho_{\mathrm{core}}(r)= 
    \begin{cases}
         \rho_{\mathrm{E}}(r_c) , & \text{if $r \leq r_c$}\\
         \rho_{\mathrm{E}}(r) , & \text{if $r > r_c$}\\
\end{cases} , 
\end{equation}
where $r_c$ denotes the core radius with typical values around 1\,kpc.

The DM profile directly impacts the overall strength of the gamma ray signal from WIMPs at the GC via the so called J-factor, the integral over the line of sight ($l$) and the solid angle ($\Omega$) of the DM density $\rho$. For the case of WIMP annihilation:
\begin{equation}
    J = \int_{\Delta\Omega} d\Omega \int dl \rho^2, 
\end{equation}
and the photon flux is proportional to $J$, with its overall normalization determined by the velocity-averaged annihilation cross-section $\sigma v\, \simeq 3\times 10^{-26} \text{cm}^3 \text{s}^{-1} $. In ~\citep{DJUVSLAND2023}, we normalised our calculations with $J = 1.53 \times 10^{22}\,\text{GeV}^2 / \mathrm{cm}^5$ , the J-factor for an NFW profile integrated over the full sky \citep{FermiJ}. 

Our previous work also showed that energetic electrons ($>\,10\,$GeV - as expected from heavy WIMP signals) cool locally in the GC region. This is because their cooling timescales are shorter than the timescales of their propagation. Using the software package GAMERA~\citep{Hahn_2016}, we computed the relevant cooling timescales and secondary radiation processes. The main mechanism of electron cooling at distances from the GC that are typically studied with gamma ray telescopes is via IC emission. Therefore, the electrons emitted by a hypothetical WIMP give rise to gamma rays that add to the prompt gamma-ray signal of the WIMP. This additional radiation adds a low-energy shoulder to the prompt signal, and together with synchrotron emission, these secondary radiation processes account for up to 45\% of the energy radiated in photons (see the discussion on Fig. \ref{fig:relativeRadiationOutput} and supplementary material) of the WIMP (depending on the mass of the WIMP and its annihilation channel).

%% file: wimps.tex
\begin{figure}[H]
    \centering
    \includegraphics[width=0.85\linewidth]{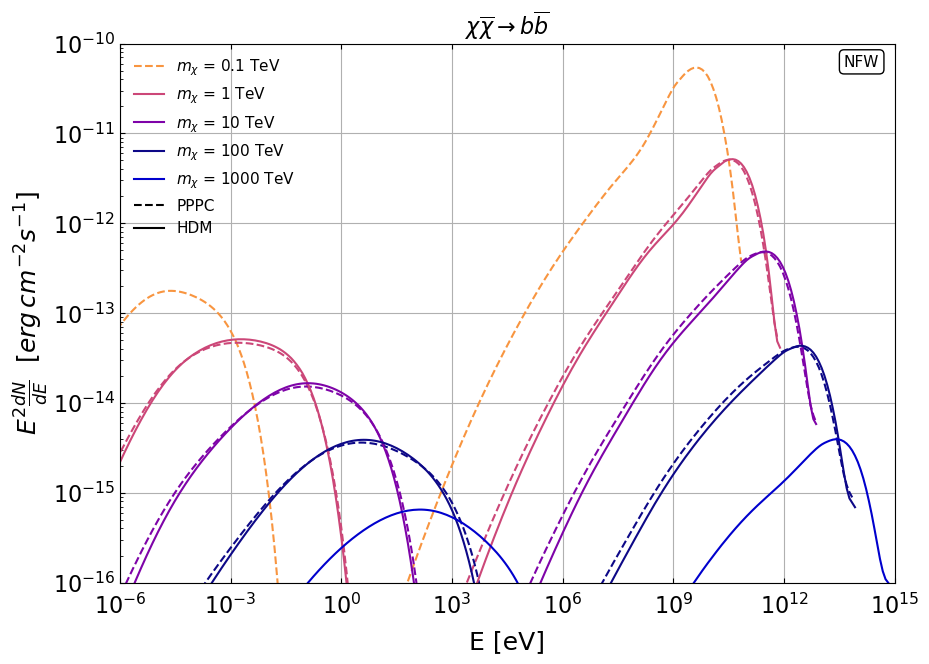}
    \caption{Total photon spectrum ( prompt $\gamma $ + IC + synchrotron emission) from WIMP annihilation to $b\overline{b}$ pairs at the GC for different WIMP masses. The solid lines show the spectra obtained with inputs from HDM~\citep{HDM}, whereas the dashed lines are computed with inputs from PPPC~\citep{Cirelli_2011}. Electrons are exposed to radiation and magnetic fields, calculated at a distance of 100 pc from the GC, and the spectra are normalized with the thermal cross section $\sigma v\, \simeq 3\times 10^{-26} \text{cm}^3 \text{s}^{-1} $,  and $J = 1.53 \times 10^{22}\,\text{GeV}^2 \text{cm}^{-5}$.}
    
    \label{fig:HDMvsPPPC}
\end{figure}

For our current study, we employ the energy spectra of the products of WIMP annihilation, i.e. the prompt photon or electron spectra, from~\citep{HDM} referred to as HDM in the following. In contrast to the computations we used in our previous studies (PPPC from~\citep{Cirelli_2011, Ciafaloni_2011}), these spectra are provided for a larger range of WIMP masses and include all relevant electroweak interactions. We used the new input spectra and revisited our previous work in \citep{DJUVSLAND2023}, where we modelled in situ cooling of the energetic electrons produced in WIMP annihilations. The electron spectrum for a given WIMP mass and annihilation channel was subjected to GC radiation and magnetic fields by assuming field strengths as evaluated at a fixed distance from the GC (where we chose 100\,pc as the baseline, as this corresponds to the typical region of interest for gamma-ray WIMP searches). As mentioned previously in Section~\ref{sec:gc}, the radiation output was normalised according to the J factor integrated over the whole sky.
We compare the output from the new spectra (HDM) to our original results (using PPPC), shown in Figure~\ref{fig:HDMvsPPPC}. 
Fair agreement is found between the total photon spectra obtained with the different inputs. The total spectrum represents the sum of the prompt and secondary emission, with synchrotron emission dominating at the lowest energies and IC adding a low-energy shoulder to the prompt emission at the highest energies. The observed decrease in signal strength with increasing WIMP mass is expected, as the photon flux is proportional to the inverse of the squared WIMP mass. % the decrease seen isn't strictly inverse squared, although that is the major effect. There is also a competing increase of particle production for higher masses.

 %Maybe we can make it explicit here that left bump shows synchrotron, right includes prompt with a lower energy IC tail. Also, this J factor is of Fermi NFW 1.53x10^22

The PPPC computations provide the input spectra for DM masses ranging from 5\,GeV to 100\,TeV, while the HDM-spectra are available for WIMP masses between 500\,GeV to 10$^{19}$\,GeV (in case of WIMP annihilation). Therefore, we only show one curve in Figure~\ref{fig:HDMvsPPPC} for the lowest and highest WIMP masses, 0.1\,TeV and 1000\,TeV respectively. 
We do not consider WIMP masses above 1000\,TeV as very heavy thermal WIMPs are violating unitarity constraints (see e.g.~\citep{PhysRevLett.64.615}). Yet, we want to point out, that there are models that overcome this challenge (see e.g.~\citep{Harigaya_2016}). Furthermore, we note the possibility of decaying WIMPs, which can yield particle spectra identical to those produced by annihilating WIMPs with half the mass of the decaying particle (see e.g.~\citep{Cirelli_2011} or \citep{Ciafaloni_2011}). 

\begin{figure}[H]
    \centering
    \includegraphics[width=0.99\linewidth]{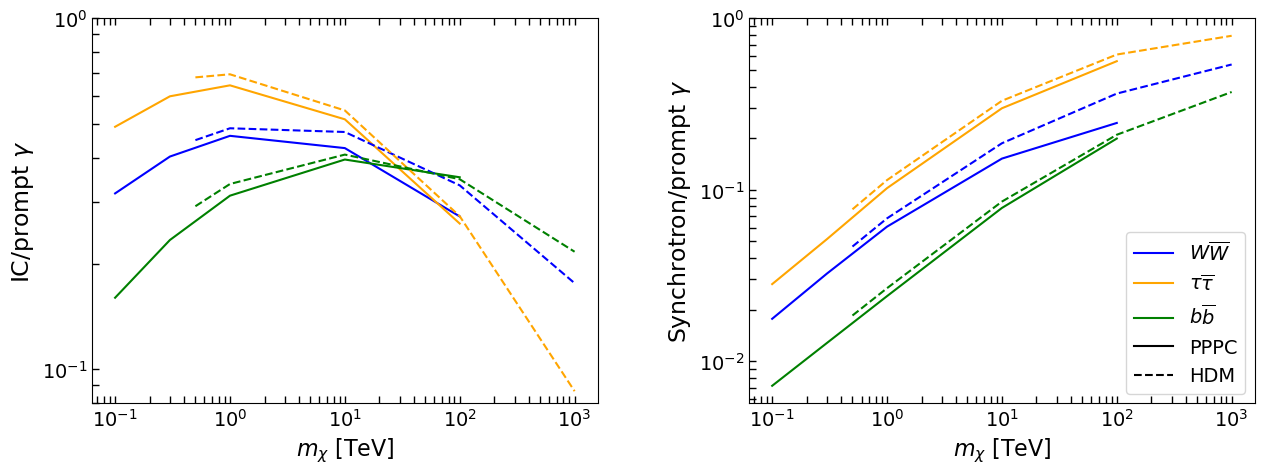}
    \caption{Ratio of the energy radiated in secondary and prompt emission of Galactic WIMP annihilation as a function of the WIMP mass m$_\chi$. The colours indicate the annihilation channel, while the line style indicates the source of the particle spectra. %The left panel shows the ratio of inverse Compton to prompt photon emission while the right panel shows the ratio of synchrotron and prompt photon emission.}
    }
    \label{fig:relativeRadiationOutput}
\end{figure}

With the wider range of WIMP masses available, we study the relative radiation output of IC and synchrotron emission with respect to the prompt photon emission as a function of the WIMP mass (see Figure~\ref{fig:relativeRadiationOutput}). 
The radiation power output is calculated by integrating $E \times dN/dE$ over the full energy range. Again, a fair agreement between the ratios obtained with the different inputs is observed for WIMP annihilations into $\tau\overline{\tau}$ and $b\overline{b}$ pairs while the ratios deviate in the $W\overline{W}$ channel with increasing WIMP mass. The deviation is expected, as HDM calculations, with the relevant electroweak corrections, produce less prompt photons than those from PPPC at the highest energies for heavier WIMPs annihilating in the $W\overline{W}$ channel. A visual comparison of the prompt photon spectra between PPPC and HDM calculations for the $W\overline{W}$ annihilation channel confirms the same. Although the difference in photon counts is small, these are the highest-energy photons. This results in comparatively larger radiated energy values from the prompt photons in PPPC, while the energy spectra of electrons remain almost identical, producing similar radiated energy from secondary emission. Comparing the energy output of IC to prompt $\gamma$ emission then produces ratios which are slightly smaller for the $W\overline{W}$ channel for HDM produced spectra. Overall, the ratio of IC to prompt emission increases in the GeV-regime and drops off again for higher WIMP masses. 

This drop can be attributed to the Klein-Nishina effect, where the IC scattering cross-section decreases with increasing photon energy. The ratio of synchrotron and prompt emission, on the other hand, increases steadily without suppression for higher WIMP masses. We can attribute this to a larger production of high-energy electrons from heavier WIMPs, relative to their prompt $\gamma$ emission. We stress that the trends shown in Fig. \ref{fig:relativeRadiationOutput} only highlight the relative importance of the energy radiated in secondary emission for the three annihilation channels. The expected flux spectra of the different emission processes are consistent with the energy content and distribution of direct photons and electrons, both of which are dictated by the annihilation channel and mass of WIMP. Secondary emission processes are especially relevant
for leptonic channels ( such as $e\overline{e},\, \mu\overline{\mu}, \,\tau\overline{\tau},$ etc. ), where electrons carry more energy from parent WIMPs. For instance, for $m_{\chi} =1\,\text{TeV}$, $\sim 0.30\,\text{TeV}$ energy is contained in prompt photons while electrons hold $\sim 0.31 \,\text{TeV}$ for  $\tau\overline{\tau}$  channel. Fig. \ref{fig:relativeRadiationOutput} confirms that secondary emission is relatively important for $\tau\overline{\tau}$ channel, and a quick observation gives secondary emission to account for $\sim 45 \%$  (\begin{math}
    \text{IC} + \text{Synchtrotron}/(\text{prompt}\, \gamma  + \text{IC} + \text{Synchrotron}) \sim 0.8/(1 + 0.7 + 0.1)
\end{math}) of the total radiation output for that channel. For $W\overline{W}$, prompt photons contain $ 0.41 \,\text{TeV}$ and electrons $0.37\,\text{TeV}$. Thus, although $W\overline{W}$ channel exhibits stronger emission from both direct and secondary photons, its ratios lie below those of $\tau\overline{\tau}$  channel. The ratios for $b\overline{b}$ are similarly consistent with its prompt photons containing $\sim 0.50 \,\text{TeV}$ and electrons $\sim 0.32\, \text{TeV}$. %Factors such as the choice of DM profile, modelling of external radiation and magnetic fields, transport of charged particles, etc. also affect the expected spectra, which will be discussed in the next sections.

 %gamma-gamma interaction ?

%% file: approach.tex
The DM mass inside the solar circle halo (radius $\sim$ 8\,\text{kpc} from the GC), considering an NFW profile, is around $3.9\times 10^{10}$ solar masses ~\citep{Sofue2011}. We found an $8\,\text{kpc}^3$ cube centred at the GC (occupying $\sim$ 0.37 per cent of that halo volume, and subtending a maximum angle of $\sim 12$ degrees from the GC) to contain $\sim 10^9$ solar masses. This significant amount of DM, although contained in a very small volume, still resides in varying external environments near the chaotic GC. In this work, we consider a more realistic secondary emission scenario by accounting for the 3D distributions of WIMP DM, radiation and magnetic fields and the transport of direct electrons via wind and diffusion within this $8 \,\text{kpc}^3$ volume. Thus, the electrons produced at one location can be transported and influenced by the external conditions at other locations, leading to secondary emissions whose detailed characteristics may differ from those described in our previous work. For this purpose, we employ the methodology used in~\citet{Depaoli25} and modify the software to our needs. The framework of~\citet{Depaoli25} was originally developed to model and investigate the origin of the Galactic radio break~\citep{Orlando2018}. It simulates particle diffusion and advective transport driven by galactic winds. It also evaluates particle radiation (with GAMERA) through synchrotron and IC processes and is used to study the diffuse emission of the Galaxy across different wavelengths, from radio to high energies, accounting for variations in magnetic and radiation fields with height above the Galactic plane. 

\subsection{Setup}\label{subsec:setup}

%3D fields (show plots?)
Considering the inner $2\times2\times2 \,\text{kpc}^3$ region near the GC, the volume is modelled by a 3D Cartesian grid. We divide our volume into a number of columns, each being a complete simulation unit of electron injection, transport, cooling and radiation. For our case, we consider 121 columns perpendicular to the Galactic plane. The centre of the central column coincides with the GC ($x$ = $y$ = $z$ = 0\,pc). Each column represents a volume of $200\,\text{pc} \times 200\,\text{pc} \times 2000\,\text{pc}$. The columns are further divided along the $z$-axis into 20 'spatial bins', and the radiation and magnetic fields are evaluated at the bin centres. These field values are taken into account while evaluating radiative cooling processes occurring in such a bin. We have a total of $121\times20 = 2420$ spatial bins for the volume considered. The model in~\citep{Depaoli25} assumes that vertical transport along the z-axis (perpendicular to the $x$-$y$ Galactic plane) is the dominant mechanism and that lateral transport in the $x$- and $y$-directions is negligible. Therefore, particle transfer between adjacent columns can be ignored. A symmetric wind profile, such as a linearly increasing wind from 0$\,\text{km/s}$ at $z=0$ to 200$\,\text{km/s}$ at $|z|=1 \,\text{kpc}$, is implemented. We further employ the same Kolmogorov diffusion formalism used in our previous work. The specifics of the WIMP model, such as WIMP mass, annihilation channel, DM density profile, etc, are initialized.

\subsection{Injection, Transport, Cooling, and Radiation}\label{subsec:simulation}

Electrons are injected in a column following the square of the specified DM density distribution along the column. In our work we mostly focus on cored Einasto distributions with core radii of 10\,\text{pc}, 100\,\text{pc} and 1\,\text{kpc}, and adapt the commonly used parameters $\rho_0 = 0.081$\,GeV/cm$^3$, $\alpha = 0.17$ and $r_s = 20$\,kpc (Eq. \ref{eq: einasto}, \ref{eq: cases}). The normalisation of the electrons according to the $dN/dE$ energy spectrum (provided by HDM or PPPC) is accounted for in the simulation. During the simulation, electrons undergo both advection driven by the Galactic wind field and diffusion.
At each time step, electrons contained in a spatial bin cool radiatively by synchrotron and IC radiation and these emission values are stored. We ignore Bremsstrahlung radiation, as it appeared to be relatively insignificant in our previous work, especially for the case of heavy WIMPs. Electrons also lose energy through adiabatic cooling due to variations in the velocity field. To ensure numerical stability and accuracy, the time step $\Delta t$ is dynamically adjusted based on the timescale of the most restrictive process among cooling, advection and diffusion. The simulation stops when the maximum synchrotron frequency of electrons, as evaluated in ~\citep{TeraelectronvoltAstronomy}, falls below a threshold, here assumed to be 10~MHz.
The simulation also stops if all particles have left the simulation volume. The approach ensures that the final electron and photon spectrum calculated over the sum of all the time steps reflects an equilibrium situation of the continuous electron injection, transport, cooling and radiation expected to be occurring at the GC.

A Cartesian grid of the secondary radiation luminosities is constructed using the computed luminosities across the 3D distribution of spatial bins.
The final radiation output produced in a column is equal to the sum of the radiation luminosities in all the spatial bins and is normalized to the dark matter annihilation content in the column. This weighting varies for different columns and depends on the distance of the columns from the Galactic centre - the closer to the GC, the greater the annihilation content. Together with the spatial distribution along the $z$ axis, our approach incorporates the 3D distribution of DM. The total secondary radiation output 
%%in $\text{TeV}\, \text{cm}^{-2} \,\text{s}^{-1}$ (or in $\text{erg}\, \text{cm}^{-2} \,\text{s}^{-1}$, $1\,\text{TeV} = 1.602 \,\text{erg}$) 
from our volume is approximated by the weighted sum of the luminosities of the columns divided by $4\pi d^2$, where $d=8.3 \,\text{kpc}$. Finally, the prompt $\gamma$ spectrum is normalized and added to the secondary radiation output, taking the rate of total annihilation in the volume, the result of which is similar to the one obtained from considering the J factor over an aperture angle of 12$^{\circ}$ ($\Delta \Omega \simeq 0.13$ sr).
 \par

%% file: results.tex
\begin{figure}[H]
    \centering
    \includegraphics[width=0.9\linewidth]{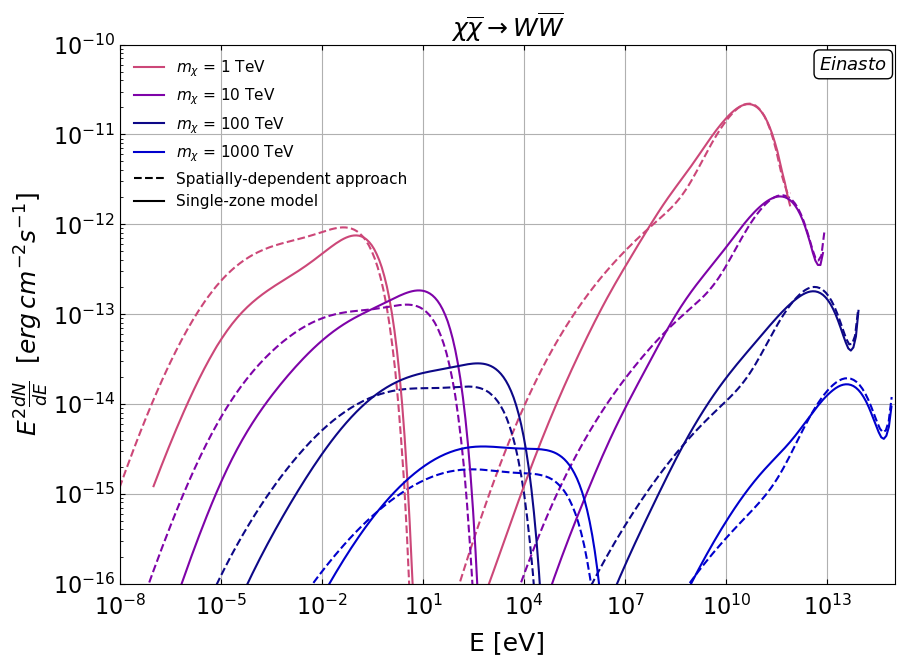}
    \caption{Comparison of the total photon spectra from WIMP annihilation to $W$ bosons computed with the spatially-dependent approach (dashed lines) and single-zone model (solid lines). The thermal cross section is set to $\sigma v = 3\times 10^{-26} \,\text{cm}^3\text{s}^{-1}$, and the resultant spectra for the spatially-dependent approach (this work) takes into account dark matter contained within an $\sim 8 \,\text{kpc}^3$ cube centred at the GC with an Einasto distribution.
    To enable direct comparison of the approach presented in \citep{DJUVSLAND2023}, the single-zone results are normalized with a J-factor of $\sim 8.15\times10^{22} \,\text{GeV}^2 \text{cm}^{-5} -$ the value obtained for an \textit{Einasto} distribution when integrated over $\theta \sim 12^\circ$ which contains the $\sim 8 \,\text{kpc}^3$ cube.}
    \label{fig:3Dto1Dcomp}
\end{figure}

The results of the spatially-dependent model are greatly aligned with the results we obtain with the single-zone model employed previously, although not an exact match. Figure~\ref{fig:3Dto1Dcomp} shows the comparison of the results of the two models for different WIMP masses in the $W\overline{W
}$ annihilation channel. Here we reproduced the results of~\citep{DJUVSLAND2023} with a cuspy Einasto (10 $\text{pc}$ core radius) normalisation, which we refer to as $Einasto$ hereafter. Good agreement is observed at the highest energies, where the direct photon component dominates. A reasonable match is found at medium and low energies, which are dominated by IC and synchrotron radiation, respectively. In general, we observe the single-zone approach to produce more secondary radiation at higher energies, while the spatially-dependent approach emphasizes the lower energy tails of the two secondary radiations. This is expected, as in the single-zone model, the electrons experience fixed, high radiation and magnetic field strengths as present very close (at $r=100 \,\text{pc}$) to the GC, resulting in stronger radiation at high electron energies.

\begin{figure}[H]
    \centering
    \begin{subfigure}[b]{0.48\textwidth}
        \centering
        \includegraphics[width=\linewidth]{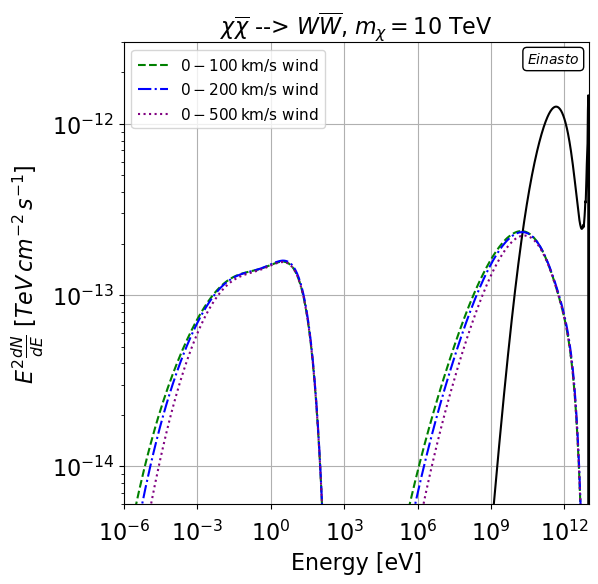}
        \caption{}
        \label{fig:wind}
    \end{subfigure}
    \hfill
    \begin{subfigure}[b]{0.48\textwidth}
        \centering
        \includegraphics[width=\linewidth]{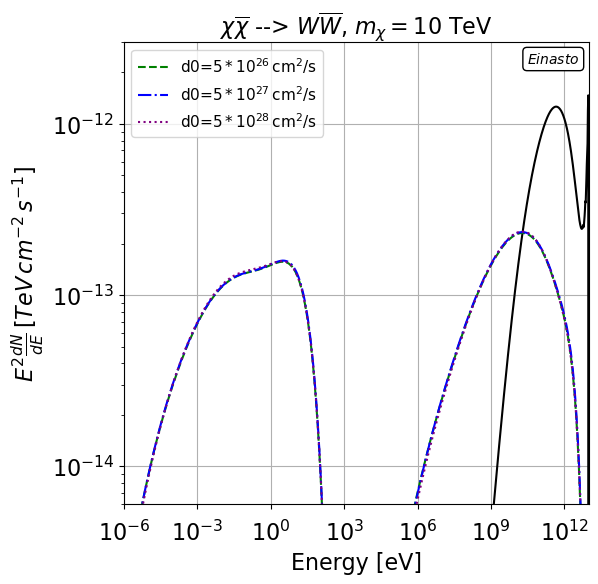}
        \caption{}
        \label{fig: diffusion}
    \end{subfigure}
    \caption{Robustness of the spatially-dependent model under varying particle transport conditions. The left panel shows the photon spectrum for different wind strengths, while the right panel shows the spectrum for three different diffusion coefficients. The prompt $\gamma$ component, shown in black, remains unaffected.}
    \label{fig:wind+diffusion}
\end{figure}

The robustness of the spatially-dependent model with respect to different conditions of particle transport is shown in Figure~\ref{fig:wind+diffusion}.  We consider results for a particular case of WIMP mass (10 TeV) and a channel of annihilation ($W\overline{W}$), with the same set of normalizations. The left panel shows the secondary emission for three different GC wind strengths. In each case the wind strength increases linearly from 0\,km/s at $z = 0$\,pc to the value v at $z = 1$\,kpc, where v is set to 100\,km/s, 200\,km/s and 500\,km/s respectively. The increasing wind speeds only slightly affect the secondary emission, where less radiation is expected, the stronger the winds are. This is due to the fact, that for strong winds, the lower energy electrons are transported out of the region before they cool and emit photons.   

The right panel of Figure~\ref{fig:wind+diffusion} shows the photon spectrum for different diffusion coefficients $d_0$. The spectrum remains almost unchanged for the different coefficients. %, pointing to the fact, that the particle transport at the GC is dominated by the advection winds.

\begin{figure}[H]
    \centering
    \begin{subfigure}[b]{0.48\textwidth}
        \centering
        \includegraphics[width=\linewidth]{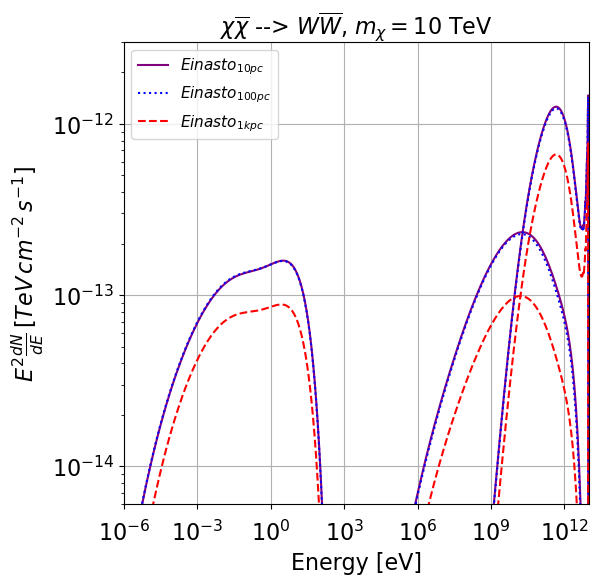}
        \caption{}
        \label{fig:profiles}
    \end{subfigure}
    \hfill
    \begin{subfigure}[b]{0.48\textwidth}
        \centering
        \includegraphics[width=\linewidth]{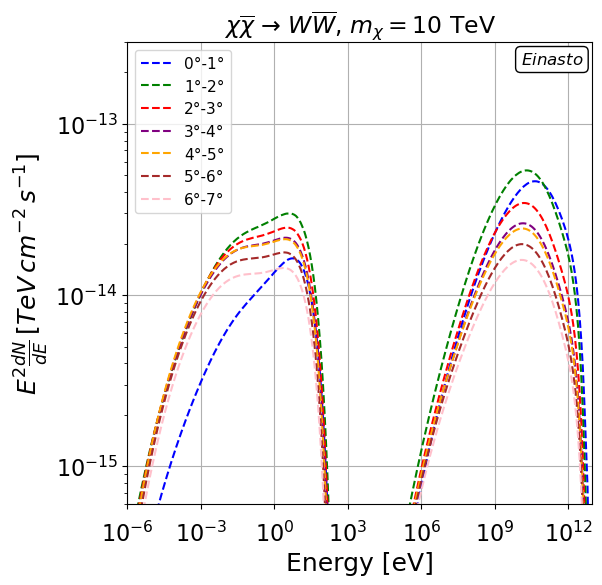}
        \caption{}        
        \label{fig:radialDist}
    \end{subfigure}
    \caption{Photon spectra from WIMP annihilation to $W$ bosons for different DM profiles (left). The variation in the secondary photon spectra as a function of radial distance from the GC (right) is shown by the dashed lines.}
\end{figure}

The impact of DM density profiles at the GC on the gamma-ray signal is as shown in Figure~\ref{fig:profiles}. For a small core radius (below 100\,pc), no big deviation from the cuspy Einasto profile is observed. For larger core radii (above 1\,kpc), the expected gamma ray flux is reduced. The core size has an overall negligible effect on the shape of the photon spectrum.

The spatially-dependent model, and specifically, the 3D distribution of the spatial bins, also allows for the study of the secondary photon emission as a function of the radial distance from the GC (Figure~\ref{fig:radialDist}). To this end, the expected flux is computed in 1$^{\circ}$ annuli around the GC by summing up the luminosities of the spatial bins contained in the annular solid angles and dividing by $4\pi d^2$. The shape of the spectra %as well as the normalisation 
is comparable for the different annuli, with the exception of the innermost one. A significant population of hardened electrons from IC-dominated cooling in the K-N regime is apparent from the dashed-blue curve synchrotron spectra~\citep{Hinton2007}. DM searches in this band are highly compromised due to strong astrophysical backgrounds. The differences in the emission also strongly depend on the DM density distribution (cuspy versus cored) and the channel of annihilation. The emission fluxes in such angular bands for all cases are provided in the supplementary material \citep{shinde_2025_16414129}.

%% file: discussion.tex
%Detectability? Limits could be set using the radio maps

%Branching fraction may or may not be 1, but the general results hold- IC adds to lower energy tail to prompt spectrum.

 We note that the resolution of the simulation setup (200 \text{pc} in $x$, $y$, 100 \text{pc} in $z$) adopted in this study was chosen to balance the computational efficiency and output detail. We verified that the key results remain consistent across different resolution settings.  Additionally, we note that there are uncertainties in the radiation field, including anisotropies in the ISRF near the GC that can alter the morphology and normalization of the computed IC emission. However, larger uncertainties on the relative fluxes in synchrotron and IC components are expected to be introduced by the less constrained modelling of the magnetic fields. We tested our model under different scenarios of external magnetic and radiation field strengths, assessing the impact on the two secondary emission components. We also tested the scenario of a much higher B field ( $\sim 100\, \mu\text{G}$) within the central few hundred parsecs of the GC. The resulting spectra exhibit the expected reduction in IC emission as synchrotron cooling becomes efficient. The complete emission spectra (including ready-to-use secondary emission fluxes for different angular bands around the GC, similar to Fig. \ref{fig:radialDist}, and the total output including the prompt emission) for all three annihilation channels and WIMP masses from 1 to 1000 TeV, along with these validation checks, are provided in the supplementary materials. 

 %Mentioning about gamma -gamma attenuation here
Together with the simplifying assumptions already discussed, we note that we have not taken into account gamma-ray attenuation by processes such as photon-photon interaction (\begin{math}
    \gamma +\gamma \rightarrow e^+ + e^-)
\end{math}, with target photons provided by radiation fields in interstellar and intergalactic space. For photon energies and trajectories relevant to this work, gamma-ray attenuation due to thermal emissions from dust peaks at photon energy $E_{\gamma} \simeq 150 \,\text{TeV}$ with gamma-ray survival probability $P_{\text{surv}} \gtrsim 0.8$, and increases towards a larger absorption maximum for $E_{\gamma} \simeq 2.2\, \text{PeV}$ ($P_{\text{surv}}\simeq 0.3$) due to interaction with the CMB \citep{gammagamma2016}. The later strongly affects the highest energy prompt photons for the extremely massive WIMP cases like $m_{\chi} = 1000\, \text{TeV}$. For photons (prompt or IC) in the range of 10s to 100s TeV, the attenuation effect, although small but non-negligible, is a less severe concern compared to the astrophysical backgrounds along the GC line-of-sight. Absorption from thermal emission from dust also decreases rapidly with increase in the galactic latitude $b$, with \citep{gammagamma2016} showing $\gtrsim 90\%$ photons at $150 \,\text{TeV}$ surviving when sources are above the GC with $b=5^\circ$. Moreover, even considering the WIMPs with $m_{\chi}=1000 \,\text{TeV}$, the  $E^2\frac{dN}{dE}$ signal peaks much below $1000 \,\text{TeV}$, between $\sim 10$ to $\sim 100 \,\text{TeV}$. For these reasons, gamma-ray searches for $\sim 100 \,\text{TeV} $ WIMPs avoiding the region close to the GC line-of-sight and the galactic plane are still worthwhile, with the lower energy IC photons suffering an even lesser suppression effect from $\gamma - \gamma $ interaction. A detailed examination of the impact of photon-photon absorption on WIMP spectra according to energy and varying emission distances around the GC can be produced, but this is beyond the scope of this work.

More importantly, we verified and improved upon the findings of our previous work - inverse Compton emission is significant and adds a lower energy shoulder to the prompt gamma-ray spectrum of annihilating WIMPs. Our results come just in time for the preparation of the data analyses for the future observatories in the southern sky - CTAO and SWGO, which are expected to deliver a leap in sensitivity to high-energy gamma rays (up to $\sim 100$ TeVs) from the GC. These instruments may offer the opportunity to probe the heaviest WIMPs, whose signals could be enhanced by inverse-Compton emission. CTAO is expected to reach the thermal cross section for TeV-scale DM \citep{Acharyya_2021}, while additionally advancing the measurement of the diffuse interstellar emission.  SWGO will further constrain a thermal DM signal in the 100 TeV range. %These sensitivity limits are, however, derived by accounting only for the prompt gamma ray emission.
 DM searches from both CTAO and SWGO will be highly complementary, and joint observations could result in great statistical significance by coincidence detection \citep{SWGO_science_2025}. A thermal cross section for WIMP masses from a few GeV to $\sim 80 \,\text{TeV}$ could be probed \citep{Viana_2019} from the combined sensitivity of SWGO, CTAO and Fermi-LAT, especially for channels such as $\tau\overline{\tau}$ and $W\overline{W}$.
 The inclusion of secondary IC photons, aided by their angular distribution, could help put stronger limits on the WIMP parameter space. %, and could increase the discovery potential of both facilities if a signal is observed. 
 At radio wavelengths, the expected WIMP-induced synchrotron signal is much lower than the observed synchrotron emission from the GC (see supplementary material). This confirms that radio detection prospects remain poor.

%% file: conclusions.tex
In order to fully test the WIMP paradigm, it is too simplistic to only consider the primary gamma ray signal coming directly from WIMP annihilation. Indirect photon emission also contributes sizeably, as we showed in this study where we considered the GC as the target. We developed a more sophisticated 3-dimensional model that confirms our initial findings: IC component is non-negligible over a wide range of WIMP masses.

We compared two different computations of the energy spectra of the WIMP products (HDM and PPPC) and found negligible impact on our results. Our results are also robust under the change of the transport parameters (wind strength and diffusion coefficients) and do not strongly depend on the WIMP distribution.

All in all, our study confirms the importance of the indirect photon component of the gamma ray WIMP signal from the GC. It should therefore be included in future WIMP searches. Through its alteration of the photon spectrum, this component can be crucial in determining the exact particle nature of dark matter if a WIMP signal is observed.